\documentclass[preprint,3p,times,twocolumn]{elsarticle}

 \usepackage{graphicx}

\usepackage{amssymb}
\usepackage{bm}

\usepackage{amsmath}
\journal{Sci China Ser G-Phys Mech Astron}
\usepackage{color}

\begin{document}

\begin{frontmatter}



\title{Long-range adiabatic quantum state transfer through a linear array of
quantum dots}


\author[sd]{CHEN~Bing\corref{cor}}
\ead{chenbingphys@gmail.com}
\author[sd]{FAN~Wei}
\author[sd,sg]{XU~Yan\corref{cor}}
\ead{x1y5@hotmail.com}
\cortext[cor]{Corresponding author}

\address[sd]{College of Science, Shandong University of Science and Technology, Qingdao
266510, China}
\address[sg]{Centre for Quantum Technologies, National University of Singapore, 3 Science drive 2, Singapore 117543}

\begin{abstract}
We introduce an adiabatic long-range quantum communication proposal based on a
quantum dot array. By adiabatically varying the external gate voltage
applied on the system, the quantum information encoded in the electron can
be transported from one end dot to another. We numerically solve the schr%
\"{o}dinger equation for a system with a given number of quantum dots. It is
shown that this scheme is a simple and efficient protocol to coherently
manipulate the population transfer under suitable gate pulses. The
dependence of the energy gap and the transfer time on system parameters is
analyzed and shown numerically. We also investigate the adiabatic passage in
a more realistic system in the presence of inevitable fabrication
imperfections. This method provides guidance for future realizations of
adiabatic quantum state transfer in experiments.
\end{abstract}

\begin{keyword}
adiabatic passage \sep tight-binding model \sep quantum state transfer

PACS: 03.65.-w, 03.67.Hk, 73.23.Hk
\end{keyword}

\end{frontmatter}


\section{Introduction}
\label{sec:Intro}

In quantum information science, quantum state transfer (QST), as the name
suggests, refers to the transfer of an arbitrary quantum state $\alpha
\left\vert 0\right\rangle +\beta \left\vert 1\right\rangle $ from one qubit
to another. There are two major mechanisms for QST in quantum mechanics. The first
approach is usually characterized by preparing the quantum channel with
an \textit{always-on} interaction where QST is equivalent to the time
evolution of the quantum state in the time-independent Hamiltonian~\cite%
{Bose1,Song,Christandle1}. However, these approaches require precise control
of distance and timing. Any deviation may lead to significant errors. The
other approach has paid much attention to adiabatic passage for coherent
QST in time-evolving quantum systems, which is a powerful tool for
manipulating a quantum system from an initial state to a target state. This
method of population transfer has the important property of being robust
against small variations of the Hamiltonian and the transport time, which is
crucial experimentally since the system parameters are often hard to control.
Recently,
the adiabatic method has been applied to a variety of physical systems to
realize coherent QST. Among these, the typical scheme for coherently spatial
population transfer has been independently proposed for neutral atoms in
optical traps~\cite{Eckert} and for electrons in quantum dot (QD) systems%
~\cite{CTAP} via a dark state of the system, which is termed coherent
tunneling via adiabatic passage (CTAP) following~\cite{CTAP}. In such a
scheme, the tunneling interaction between adjacent quantum units is
dynamically tuned by changing either the distance or the height of the
neighboring potential wells following a counterintuitive scheme, which is a
solid-state analog of the well-known stimulated Raman adiabatic passage
(STIRAP) protocol~\cite{STIRAP} of quantum optics. Since then, the CTAP
technique has been proposed in a variety of physical systems for
transporting single atoms~\cite{atom1,atom2}, spin states~\cite{spin},
electrons~\cite{electron1,electron2} and Bose-Einstein condensates~\cite%
{BEC1,BEC2,BEC3}. It has also been proposed as a crucial element in the
scale up to large quantum processors~\cite{LR1,LR2}.

\begin{figure}[tbp]
\center
\includegraphics[bb=138 376 416 506, width=7 cm, clip]{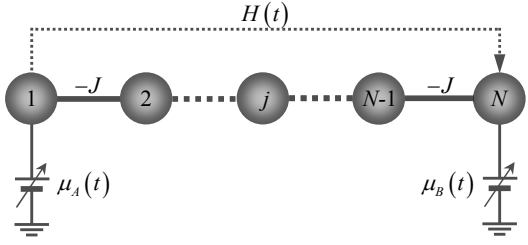}
\caption{Schematic illustrations of adiabatic QST in a multi-dot array. The
system is controlled by gate voltages, $\protect\mu _{\protect\alpha }(t)$ $(%
\protect\alpha =A,$ $B)$. By adiabatically varying the gate voltages, one
can achieve long-range QST from the left end to the right QD of the proposal.}
\label{fig1}
\end{figure}

Recently, Ref.~\cite{chen} presented a scheme to adiabatically transfer an
electron from the left end to the right end of a three dot chain using the
ground state of the system. This technique is a copy of the frequency
chirping method~\cite{CF1,CF2}, which is used in quantum optics to transfer
the population of a three-level atom of the Lambda configuration. The scheme~%
\cite{chen} is presented as an alternative to a well known transfer scheme
(CTAP)~\cite{CTAP}. However, different from the CTAP process, the protocol in
Ref.~\cite{chen} considers a three QD array with an \emph{always-on}
interaction that can be manipulated by the external gate voltage applied on
the two external dots (sender and receiver). Through maintaining the system
in the ground state, it shows that it is a high-fidelity process for a
proper choice of system parameters and also robust against experimental
parameter variations. In this paper we will consider the passage through the
$N$-site coupled QDs array (tight-binding model), which is schematically
illustrated in Fig.~\ref{fig1}. Gates applied on the two end dots control
the on-site energy of each dot. In particular, the nearest-neighbor hopping
amplitudes are set to be uniform. We first investigate the effect of system
parameters on the minimum energy gap between the ground state and the first
excited state. Taking a 5-dot structure as an example, we show that the
electron can be robustly transported from one end of the chain to the other
by slowly varying the gate voltages. This structure is easy to extend to an
arbitrary number of sites.

The paper is organized as follows. In Sec. II the model is setup and we
describe the adiabatic transfer of an electron between QDs. In Sec. III we
show numerical results that substantiate the analytical results. The last
section is the summary and discussion of the paper. 

\section{Model setup}
\label{sec:Model}

We introduce a simple tight-binding chain with uniform nearest-neighbor hopping integral $-J$ as a
quantum data bus for long-range quantum transport, see Fig.~\ref{fig1}. The
sender (Alice) and the receiver (Bob) can only control the external gate
voltages $\mu _{\alpha }(t)$ $(\alpha =A,$ $B)$, which are applied on the two
end QDs. In this proposal, the quantum information $\cos \theta \left\vert
\uparrow \right\rangle +\sin \theta \left\vert \downarrow \right\rangle $
encoded in the polarization of the electron can be transported from
Alice to Bob. In this scheme the spin state of the electron is a
conserved quantity for the Hamiltonian of the medium, so the spin state
cannot be influenced during the propagation. For simplicity, we will
disregard the electron's spin degrees of freedom in the following discussion
and just illustrate the principles of QST.

Accounting only for the occupation of the lowest single-particle state of
each dot, the system is described by an $N$-site tight-binding chain. The
Hamiltonian can be written as
\begin{eqnarray}
\mathcal{H}(t) &=&\mathcal{H}_{M}+\mathcal{H}_{C},  \notag \\
\mathcal{H}_{M} &=&-J\sum_{j=1}^{N-1}\left( a_{j}^{\dag }a_{j+1}+\text{h.c.}%
\right) ,  \label{H_t} \\
\mathcal{H}_{C} &=&\mu _{A}\left( t\right) a_{1}^{\dag }a_{1}+\mu _{B}\left(
t\right) a_{N}^{\dag }a_{N},  \notag
\end{eqnarray}%
where $a_{j}^{\dag }(a_{j})$ denotes the spinless fermion creation
(annihilation) operator at the $j$-th quantum site. $-J$ $(J>0)$is the coupling
constant, which accounts for the hopping of the electron between dots $j$ and $j+1
$. The on-site energies $\mu _{A}(t)$ and $\mu _{B}(t)$ are modulated in
Gaussian pulses to realize the adiabatic transfer, according to [shown in
Fig.~2(a)]
\begin{subequations}
\begin{align}
\mu _{A}(t)& =-\mu _{A,max}\exp \left[ -\alpha ^{2}t^{2}/2\right] ,
\label{miuL} \\
\mu _{B}(t)& =-\mu _{B,max}\exp \left[ -\alpha ^{2}\left( t-\tau \right)
^{2}/2\right] ,  \label{miuR}
\end{align}%
\end{subequations}
where $\tau $ and $\alpha $ are the total adiabatic evolution time and
standard deviation of the control pulse. For simplicity we set the two peak
values to be equal, i.e., $\mu _{A,max}=\mu _{B,max}=\mu _{0}$ $(\mu >0)$, in
the discussion that follows and choose $\mu _{0}\gg J$.

In this proposal, we will concentrate on the single-particle problem and use the ground state $\left\vert \psi _{g}(t)\right\rangle $
of the Hamiltonian $\mathcal{H}(t)$ to induce a population transfer from state $%
\left\vert 1\right\rangle $ to $\left\vert N\right\rangle $. Starting from $%
t=0$, we have $\mu _{A}(0)=-\mu _{0}$ and $\mu _{B}(0)\approx 0$. The Hamiltonian at $t=0$ reads
\begin{equation}
\mathcal{H}(t=0)=-\mu_{0} a_{1}^{\dag }a_{1}-J\sum_{j=1}^{N-1}\left( a_{j}^{\dag }a_{j+1}+\text{h.c.}%
\right).\label{EQ}
\end{equation}%
The ground state of Eq. (\ref{EQ}) is a bound state, which can be obtained via the Bethe
ansatz method. A straightforward calculation shows that
\begin{equation}
\left\vert \psi _{g}(t=0)\right\rangle  =\sqrt{1-\zeta
^{2}}\sum_{j=1}^{N}\zeta ^{j-1}\left\vert j\right\rangle ,
\label{psi_g}
\end{equation}
where $\zeta =J/\mu _{0}$. By
choosing a sufficiently large value of $\mu _{0}$, the ground state $%
\left\vert \psi _{g}(0)\right\rangle $\ can be reduced to
$\left\vert \psi _{g}(0)\right\rangle \approx \left\vert
1\right\rangle=a_{1}^{\dag }\left\vert 0\right\rangle $. To
illustrate with an example, the probability of $\left\vert
1\right\rangle $ in $\left\vert \psi _{g}(t=0)\right\rangle $ can
achieve 99.75\% when the peak voltage is set to be $\mu _{0}/J=20$.

With the same reasoning, in the time limit $t=\tau $, the parameter $\mu _{A}(t)$
goes to zero and $\mu _{B}(t)$ goes to $-\mu _{0}$. Due to the
reflection symmetry (relabelling sites from right to left) of the system, we
can see that
\begin{equation}
\left\vert \psi _{g}(t=\tau )\right\rangle =\sqrt{1-\zeta
^{2}}\sum_{j=1}^{N}\zeta ^{N-j}\left\vert j\right\rangle.
\end{equation}%
One can see that the ground state of Eq.~\eqref{H_t} evolves to be $%
\left\vert \psi _{g}(t=\tau )\right\rangle \approx \left\vert N\right\rangle $%
. Preparing the system in state $\left\vert \Psi \left( t=0\right)
\right\rangle =\left\vert 1\right\rangle \,$and adiabatially changing $\mu
_{A}(t)$ and $\mu _{B}(t)$, one can see that the system will end up in $%
\left\vert N\right\rangle $
\begin{equation}
\left\vert \Psi \left( t=0\right) \right\rangle =\left\vert 1\right\rangle
\rightarrow \left\vert \Psi \left( t=\tau \right) \right\rangle =\left\vert
N\right\rangle .
\end{equation}

\begin{figure}[h]
\center
\includegraphics[ bb=36 49 220 300, width=7 cm, clip]{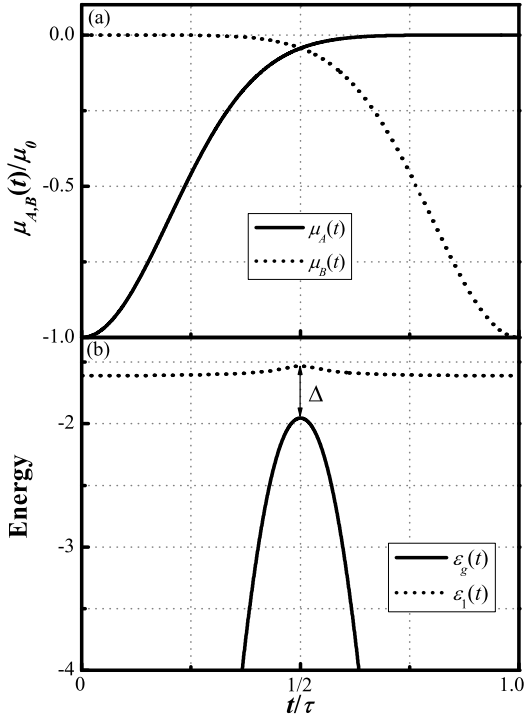}
\caption{(a) Gate voltages (in units of $\protect\mu _{0}$) as a function of
time (in units of $\protect\tau $) described in Eq.~(2). $\protect\mu %
_{A}(t) $ is the solid line and $\protect\mu _{B}(t)$ is the dashed line. (b)
The instantaneous eigenenergy of the lowest two states $\protect\psi _{1}$ and $%
\protect\psi _{g}$ though the gate pulse shown in (a) for the values of $%
\protect\mu _{0}=20$, $J=1.0$, and $\protect\alpha=5/\protect\tau$ in an $N=5$
structure. The gap is minimum at $t=\protect\tau /2$, $\Delta =\protect%
\varepsilon _{1}(\protect\tau /2)-\protect\varepsilon _{g}(\protect\tau /2)$.}
\label{fig2}
\end{figure}

The analysis above is based on the assumption that the adiabaticity is
satisfied. The crucial requirement for adiabatic evolution is
\begin{equation}
\left\vert \varepsilon _{g}(t)-\varepsilon _{1}(t)\right\vert \gg \left\vert
\langle \dot{\psi}_{g}(t)|\psi _{1}(t)\rangle \right\vert ,
\end{equation}%
which greatly suppresses the quantum transition from the ground state $|\psi
_{g}(t)\rangle $ to the first-excited state $|\psi _{1}(t)\rangle $.
Firstly, one must make sure that no level crossings occur, i.e., $%
\varepsilon _{g}(t)-\varepsilon _{1}(t)<0$. To evaluate instantaneous
eigenvalues of the Hamiltonian is generally only possible numerically. In
Fig.~\ref{fig2}(b) we present the results showing the eigenenergy gap
between the instantaneous first-excited state and ground state undergoing
evolution due to modulation of the gate voltage according the pulses given in Eq.~(2)
for $\mu _{0}=20$, $J=1.0$ and $\alpha =5/\tau$. The eigenvalues shown in
this figure exhibit pronounced avoided crossing and approach nonzero minimum
$\Delta =\varepsilon _{1}(\tau /2)-\varepsilon _{g}(\tau /2)$ at $t=\tau /2$%
. This minimum energy gap plays a significant role in the transfer, because
the total evolution time $\tau $ should be large compared to $%
1/\Delta $. In this scheme, the energy gap $\Delta $ depends both on the
number of QDs and gate voltages. To study the relationship between the total
evolution time $\tau $ and system parameters is one of the important
contributions of this paper.

\begin{figure}[h]
\center
\includegraphics[bb=42 22 220 302, width=7 cm, clip]{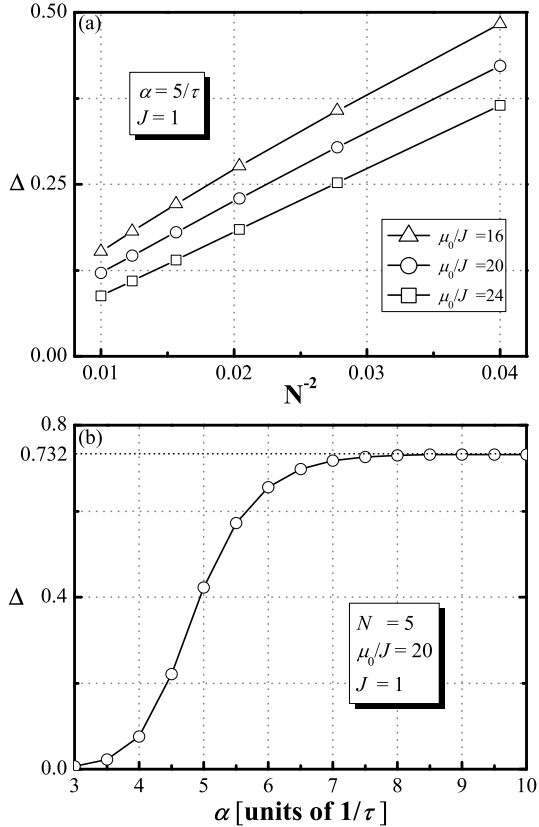}
\caption{Effect the system parameters have on the energy difference. (a) The
gap $\Delta$ obtained using a numerical method for the systems $N=5$, $6$, $7$, $%
8 $, $9$, and $10$, with $\protect\mu_{0}=16$, $20$, $24$ and $J=1.0$ are
plotted. It indicates that $\Delta \sim J^{2}/\left( \protect\mu %
_{0}N^{2}\right)$. (b) The gap $\Delta$ as a function of $\protect\alpha$. It shows the increase of the gap with increasing $\protect%
\alpha$.}
\label{fig3}
\end{figure}

Fig.~\ref{fig3} shows the effect four factors have on the energy gap $%
\Delta $. In Fig.~\ref{fig3}(a), for a given $\alpha =5/\tau$ and $J=1.0$,
we plot the energy gap $\Delta $ as a function of $N^{-2}$ for $\mu _{0}=$%
16, 20, and 24. The numerical results indicate that the gap $\Delta \sim
J^{2}/\left( \mu _{0}N^{2}\right) $, which implies that the adiabatic
transfer time of QST grows quadratically with the spatial separation of the
two end states because the minimum gap plays an opposite role for the
adiabatic QST. The other thing is that $\Delta $ is also determined by the
dimensionless parameter $\alpha$. As an example, Fig.~\ref{fig3}(b)
shows the numerically computed behavior of $\Delta $\ as a function of $%
\alpha$ with $N=5$ and $\mu _{0}/J=20$. From Fig.~\ref{fig3}(b) we see
that the gap becomes larger as $\alpha$ increases and then tends to
be a constant $0.732$, which is the gap of tight-binding chain ($N=5$)
without any on-site energy. The reason is that the bigger $\alpha \tau $ is, the
smaller the overlap amplitude of the two pulses, i.e., $\exp (-\alpha ^{2}\tau
^{2}/8)\rightarrow 0$. The energy gap $\Delta $ of $\mathcal{H}(\tau /2)$
then approaches the maximum value $2J[\cos \pi /(N+1)-\cos 2\pi /(N+1)]$.

\section{Numerical Examples}

In this section let us firstly review the transfer process of this protocol.
At $t=0$ we initialize the device so that the electron occupies site-$1$,
i.e., the total initial state is $\left\vert \Psi \left( 0\right)
\right\rangle =\left\vert 1\right\rangle $, and slowly apply gate pulses, which
results in robust transport of the electron from one end of the chain to the
other. The consequent time evolution of the state is given by the Schr\"{o}%
dinger equation (assuming $\hbar=1$)
\begin{equation}
i\frac{d}{dt}\left\vert \Psi \left( t\right) \right\rangle =\mathcal{H}%
(t)\left\vert \Psi \left( t\right) \right\rangle .  \label{SEQ}
\end{equation}

The time evolution creates a coherent superposition:
\begin{equation}
\left\vert \Psi \left( t\right) \right\rangle
=\sum_{j=1}^{N}c_{j}(t)\left\vert j\right\rangle,
\end{equation}%
where $c_{j}(t)$ denotes the time-dependent probability amplitude for the
electron to be in the $j$-th QD that obeys the normalization condition $%
\sum_{j=1}^{N}\left\vert c_{j}(t)\right\vert ^{2}=1$. At time $\tau$ the
fidelity of the initial state transferring to the dot-$N$ is defined as
\begin{equation}
F(\tau)=\left\vert \langle N\left\vert \Psi (\tau)\right\rangle \right\vert
^{2}=\left\vert c_{N}(\tau)\right\vert ^{2}.
\end{equation}

A feasible proposal should be able to perform efficient high-fidelity QST in
the shortest possible time. In order to provide the most economical choice
of parameters for reaching high transfer efficiency, we used standard
numerical methods to integrate the Schr\"{o}dinger equation for probability
amplitudes. Because the scheme relies on maintaining adiabatic conditions,
we examine the effect of system parameters on the target state population.

\begin{figure}[tbp]
\center
\includegraphics[ bb=26 8 278 537, width=7 cm, clip]{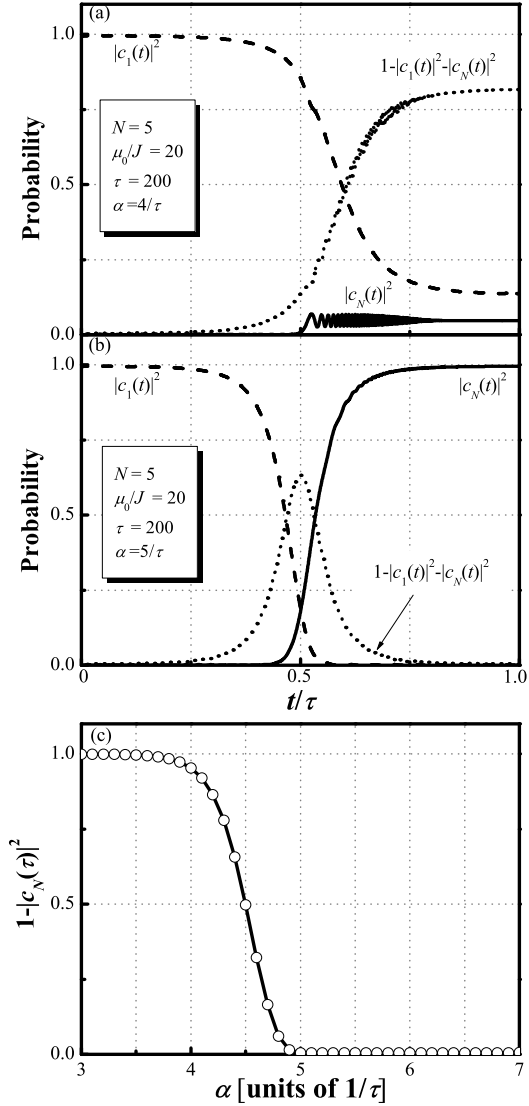}
\caption{Probability of finding a single electron in basis
states in the ground state as a function of time for the values of $N=5$, $%
J=1.0$, $\protect\mu _{0}=20$, $\protect\tau =500 $ and (a) $\protect\alpha%
=4/\protect\tau$, (b) $\protect\alpha=5/\protect\tau$. (c) Transfer fidelity $%
\left\vert c_{N}(t)\right\vert ^{2}$ as a function of
parameter $\protect\alpha$. As $\protect\alpha$ is
increased it is more able to obtain high fidelity transfer. A smaller $%
\protect\alpha$ introduces a population of excited states and
the transfer is no longer complete.}
\label{fig4}
\end{figure}

We show in Fig.~\ref{fig4} the probabilities as a function of time for
different values of $\alpha$ where we take a 5-dot structure for example.
The time behavior of $\mu _{A}(t)$ and $\mu _{B}(t)$ follow Gaussian
functions with $\mu _{0}/J=20$ [see Fig.~2(a)] and have been performed in a
finite time.

For the $\alpha =4/\tau$ case with $\tau =500$, which departs from the adiabatic
limit, we find the result in Fig.~\ref{fig4}(a). Here the population
transferred to the target state is only about 11\%. The reason is that the
population is excited to the upper energy states through nonadiabaticity. It
is necessary to point out that if one enlarged $\tau $ extremely in this
case, adiabaticity would be fulfilled, which would result in high transfer
fidelity.

On the other hand, as shown in Fig.~\ref{fig3}, enlarging $\alpha$ can
increase the level spacing between the first excited state and the ground
state and hence cause the adiabaticity of the system to become better. In Fig.~%
\ref{fig4}(b) we show the population evolution taking $\alpha =5/\tau$ as an
example and the populations of the states $\left\vert 1\right\rangle $ and $%
\left\vert N\right\rangle $ are exchanged with a fidelity of $99.5\%$.

The choice of pulse modulation is therefore important with the maximum
transfer speed ultimately controlled by the adiabatic criteria for the
transfer. To see the emergence of the adiabatic limit, we plot in Fig. 4(c)
the transfer fidelity as a function of the adiabaticity parameter $\alpha
$. One can see that as $%
\alpha $ increases there is an exponential appearance of the
adiabaticity in the ideal limit. That means the smaller the overlapping of two
pulses is, the more ideal adiabatic transfer takes place. This result is
extendible to an arbitrary number of QDs. In the discussion that follows, we
choose $\alpha=5/\tau$.

\begin{figure}[tbp]
\center
\includegraphics[bb=18 19 378 275, width=7 cm, clip]{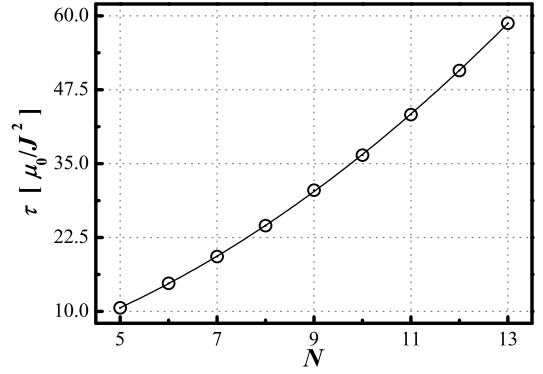}
\caption{Total evolution time $\protect\tau$ (in units of $%
\protect\mu_{0}/J^2$) as $N$ is increased under the condition that $%
\left\vert c_{N}(\protect\tau )\right\vert ^{2}\geq 0.995$. As $N$ is
increased the energy gap is decreased, resulting in longer evolution time
across the array. The solid line is the quadratic curve fitting, which
indicates that the evolution time grows quadratically with the number of
QDs.}
\label{fig5}
\end{figure}

In section II, we showed that the energy gap $\Delta $ also depends on the
system parameters, such as the number of QDs $N$ and coupling strength $J$.
In order to quantitatively determine the time needed to achieve high
fidelity QST, we solve the schr\"{o}dinger equation for constants $\mu
_{0}=20$, $J=1.0$. In Fig.~\ref{fig5} we present results showing $\tau $ as
a function of the QDs number $N$. The quadratic curve fitting shows that the
minimum possible transfer times are proportional to $N^{2}$, giving a
high-fidelity transfer of $\left\vert c_{N}(\tau )\right\vert ^{2}\geq 0.995$%
. On the other hand, the energy gap $\Delta $ decreases when the peak value
rises [see Fig.~\ref{fig3}(a)]. Consequently, the time scale $\tau $ is
proportional to and of the order of $\mu _{0}/J^{2}$ for a given $N$. To sum
up the above discussion, in practice the minimum possible transfer timescale
of this adiabatic passage will be of the same order as $N^{2}\mu _{0}/J^{2}$.

For a long enough evolution time $\tau $, the maximum fidelity of this
scheme depends on the contrast ratio between peak values $\mu _{\kappa ,max}$
$(\kappa =A,B)$ and coupling constants $J$. The reason is that small peak
values improve adiabaticity, but lead to a low fidelity because the initial
and final energy eigenstates are not the desired states $\left\vert
1\right\rangle $ and $\left\vert N\right\rangle $, respectively.

\begin{figure}[tbp]
\center
\includegraphics[bb=29 10 412 518, width=7 cm, clip]{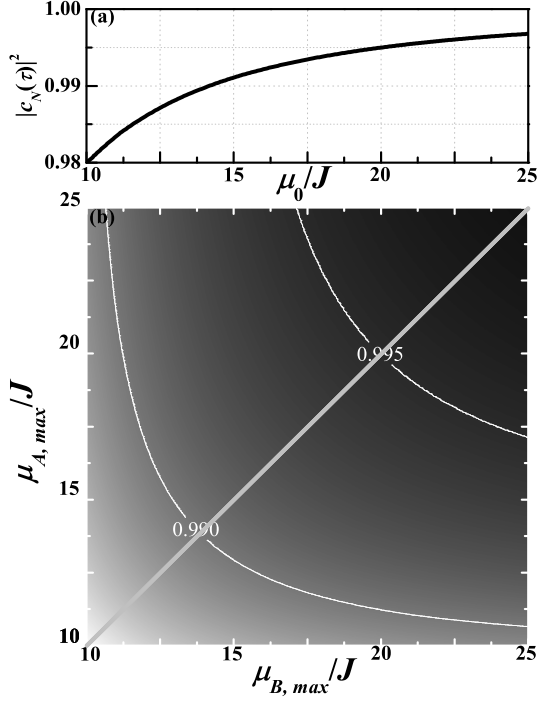}
\caption{(Color online) Plot of the transfer fidelity $|c_{N}(\protect\tau %
)|^{2}$ as a function of peak values (in units of $J$). (a) $\protect\mu %
_{A,max}=\protect\mu _{B,max}=\protect\mu _{0}$ varies from 10$J$ to 25$J$;
(b) $\protect\mu _{A,max}$ and $\protect\mu _{B,max}$ vary from 10$J$ to 25$%
J $ respectively. The contour lines, labeled with the corresponding values
of $|c_{N}(\protect\tau )|^{2}$, display the gradual increase of transfer
fidelity as $\protect\mu _{A,max}/J$ and $\protect\mu _{B,max}/J$ grow. The
fidelity is close to one ($|c_{N}(\protect\tau )|^{2}\geq 0.995$) when two
peak values are achieved for $\protect\mu _{\protect\kappa,max} \geq 20J$ $(%
\protect\kappa=A,B)$.}
\label{fig6}
\end{figure}

To determine the parameter range of $\mu _{\kappa ,max}$ $(\kappa =A,B)$
needed to achieve high fidelity transfer, we numerically integrate the
density matrix equations of motion, with varying peak values $\mu
_{A,max/J}$ and $\mu _{B,max}/J$ from 10 to 25. Fig.~\ref{fig6} shows the
transfer fidelity $|c_{N}(\tau )|^{2}$ plots as a function of $\mu
_{A,max}/J $ and $\mu _{B,max}/J$ for $\tau =1000$ and $\alpha =5/\tau$ in a
5-dot system. The fidelity approaches unity as $\mu _{A,max}$ and $\mu
_{B,max}$ increase. The figure is nearly symmetric with respect to the line $%
\mu _{A,max}=\mu _{B,max}$. Fig.~\ref{fig6}(a) is taken from Fig.~\ref{fig6}%
(b) by slicing through the diagonal gray line. We can see that to realize
near-perfect fidelity transfer ($\left\vert c_{N}(\tau )\right\vert ^{2}\geq
0.995$) one has to use peak values satisfying $\mu _{\kappa ,max}\geq 20J$ $%
(\kappa =A,B)$.

\begin{figure}[tbp]
\center
\includegraphics[bb=24 14 375 276, width=7 cm, clip]{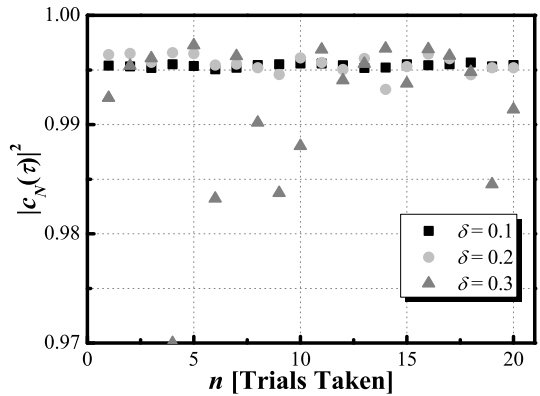}
\caption{(Color online) The transfer fidelity $\left\vert c_{N}(\protect\tau %
)\right\vert ^{2}$ for a tight-binding chain of length $N=5$ with $\protect\mu %
_{A,max}=\protect\mu _{B,max}=20J $, $\protect\tau =500$ and $\protect\alpha=5/\protect\tau$. The coupling strengths are chosen randomly from the
interval $[(1-\protect\delta )J, J ]$ for $\protect\delta =0.1$ (Square), $%
\protect\delta =0.2$ (Circle) and $\protect\delta =0.3$ (Triangle). The
number of random samples is 20.}
\label{fig7}
\end{figure}

The other advantage of this scheme is defect tolerance of the system
parameters. We now assume that the tunnel coupling has a random but constant
offset $\delta \varepsilon _{j}$, i.e. $J_{j}=J(1-\delta \varepsilon _{j})$,
where $\varepsilon _{j}$ is drawn from the standard uniform distribution on
the open interval $(0,1)$ and all $\varepsilon _{j}$ are completely
uncorrelated for all sites along the chain. We show some examples in Fig.~%
\ref{fig7} for QST in the chain of $N=5$ with maximum coupling offset bias $%
\delta =0.1,0.2$, and $0.3$. It shows that weak fluctuations (up to $\delta
=0.2$) in the coupling strengths do not deteriorate the performance of our
scheme. For $\delta =0.3$ we can see that arbitrarily perfect transfer
remains possible except for some rare realizations of $0.3\varepsilon $. To
realize high-fidelity QST transfer, the price for unprecise couplings is thus a longer transmission time.

\section{Summary}

We have introduced a robust and coherent method of long-range coherent QST
through a tight-binding chain by adiabatic passage. This scheme is realized by
modulation of gate voltages applied on the two end QDs. Under suitable gate
pulses, the electron can be transported from one end of the chain to
the other, carrying along with it the quantum information encoded in its spin.
Different from the CTAPn Scheme \cite{CTAPn}, our method is to induce
population transfer through the tight-binding chain by maintaining the system in
its ground state and this is more operable in experiments. We have studied
the adiabatic QST through the system by theoretical analysis and numerical
simulations of the ground state evolution of the tight-binding model. The result
shows that it is an efficient high-fidelity process ($\geq 99.5\%$) for a
proper choice of standard deviation $\alpha \geq 5/\tau$ and peak values $%
\mu _{0}\geq 20$ of gate voltages. For an increasing number of dots, we found
that the evolution time scale is $\tau \sim N^{2}\mu _{0}/J^{2}$. We also
consider the QST along the quantum chain if their coupling is changed by some random amount. We further find that weak fluctuations in the coupling
strength still allow high fidelity QST.

\section*{Acknowledgements}
We acknowledge the support of the NSF of China (Grant No.10847150 and No.11105086), the Shandong Provincial Natural Science Foundation (Grant No. ZR2009AM026 and BS2011DX029), and the basic scientific research project of Qingdao (Grant No.11-2-4-4-(6)-jch). Y. X. also thanks the Basic Scientific Research Business Expenses of the Central University and Open Project of Key Laboratory for Magnetism and Magnetic Materials of the Ministry of Education, Lanzhou University (Grant No. LZUMMM2011001) for financial support.

\end{document}